\documentclass[11pt]{article}
\usepackage{moriond}
\pdfoutput=1

\def\Journal#1#2#3#4{{#1} {\bf #2}, #3 (#4)}

\def\PLB{{\em Phys. Lett.}  B}

\def\EPJ{{\em Eur. Phys. J.} C}

\def\be{\begin{equation}}
\def\ee{\end{equation}}
\def\bea{\begin{eqnarray}}
\def\eea{\end{eqnarray}}

\newcommand{\Photo}{}

\begin{document}
\vspace*{4cm}
\title{Jet Results in pp and Pb-Pb Collisions at ALICE}

\author{Oliver Busch for the ALICE collaboration }

\address{Universit\"{a}t Heidelberg, Physikalisches Institut, Heidelberg, Germany}

\maketitle\abstracts{
We report results on jet production in pp and Pb-Pb collisions at the LHC from the ALICE collaboration. 
The jet cross section in pp collisions at $\sqrt{s}$=2.76~TeV is presented, as well as 
the charged particle jet production cross section and measurements of the jet fragmentation and jet shape 
in pp collisions at $\sqrt{s}$=7 TeV. NLO pQCD calculations and simulations from MC 
event generators agree well with the data. Measurements of jets with a resolution parameter 
$R$=0.2 in Pb-Pb collisions at $\sqrt{s}_{NN}$=2.76~TeV show a strong, momentum dependent 
suppression in central events with respect to pp collisions. 
The centrality dependence of the suppression of charged particle jets relative to peripheral events is presented. The 
ratio of jet spectra with $R$=0.2 and $R$=0.3 is found to be similar in pp and Pb-Pb events.
The analysis of the semi-inclusive distribution of charged particle jets recoiling from a high-$p_{\rm T}$ trigger hadron 
allows an unbiased measurement of the jet structure for larger cone radii. }

\section{Introduction}\label{sec:Intro}

Jets are collimated sprays of particles associated with hard scattered partons. The study of jet production 
and fragmentation allows us to test our understanding of perturbative and non-perturbative aspects of QCD. 
In heavy-ion collisions, jets produced in the initial stage probe the hot and dense nuclear matter 
created during the quark gluon plasma phase of the fireball evolution. Interactions with the 
medium give rise to additional induced radiation. Jet reconstruction 
aims to capture the full dynamics of jet quenching and to quantify the in-medium jet energy loss.

\section{Data analysis}\label{sec:jetReco}

Charged jets are reconstructed in the ALICE central barrel from primary charged particle 
tracks measured in the Inner Tracking System (ITS) and the Time Projection Chamber (TPC). Tracks with 
transverse momentum $p_{\rm T}>150$~MeV/$c$ in the pseudo-rapidity interval $|\eta|<0.9$ are clustered 
with the FastJet~\cite{FastJet} anti-$k_{\rm T}$ algorithm using a boost invariant $p_{\rm T}$ recombination scheme. 
We use different values of the jet resolution parameter from $R$=0.2 to $R$=0.4. Jets are selected 
so that they are fully contained within the detector acceptance. For full jet reconstruction, we include neutral particles 
(mostly neutral pions) reconstructed with the ALICE Electromagnetic Calorimeter (EMCal). Clusters in the EMCal acceptance $|\eta|<0.7$ 
and  $1.4<\varphi<\pi$ with a transverse energy $E_{\rm T}>300$~MeV are used. 
The energy of charged particles pointing to EMCal clusters is subtracted to avoid double-counting. 
The jet shape and jet constituents' transverse momentum spectra (jet fragmentation) are measured 
with charged particles in the leading (highest $p_{\rm T}$) jet in each event. The jet spectra, shape, and 
fragmentation distributions are corrected for detector effects (including the missing energy from e.g. neutrons and $K^0_L$ 
in the measurements that include the EMCal) via unfolding or bin-by-bin corrections based on detector simulations. \par 
Jet reconstruction in heavy-ion collisions proceeds against a large background from the underlying event 
uncorrelated to hard parton scattering. The average charged background energy density is evaluated  
event by event from the median $p_{\rm T}$ density of FastJet~\cite{FastJet} $k_{\rm T}$ clusters 
and subtracted jet by jet. For charged+EMCal jet reconstruction, the charged background density is scaled to the level of 
charged+neutral particles. The spectra are corrected for detector effects and background flucutations via unfolding. 

\section{Results from pp collisions}\label{sec:resultspp}

\begin{figure}[h]
\begin{minipage}{0.37\linewidth}
\hspace{1.6cm}  \centerline{\includegraphics[width=0.95\linewidth]{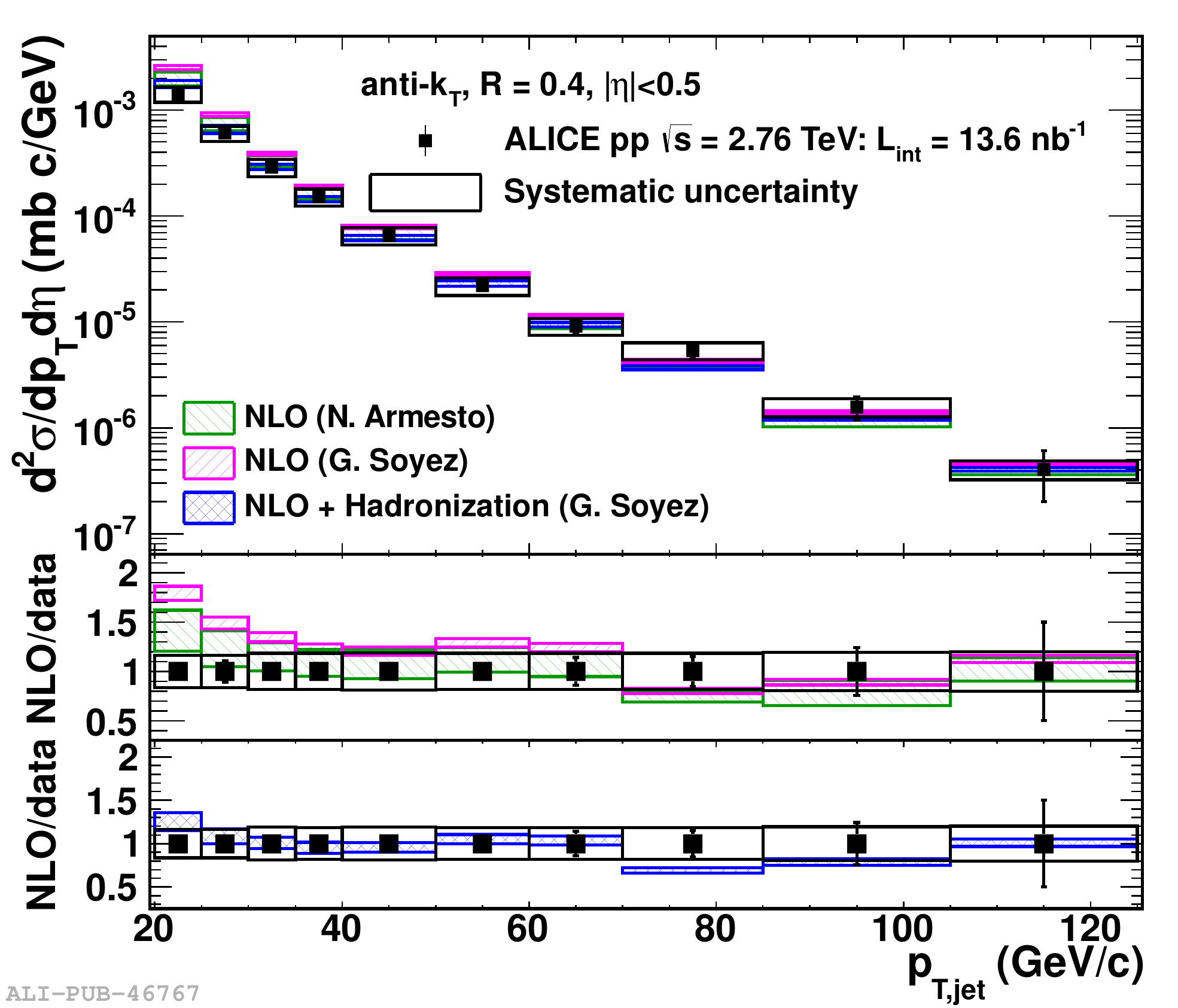}}
\end{minipage}
\begin{minipage}{0.37\linewidth}
\hspace{2.0cm} \vspace{-0.75cm} \centerline{\includegraphics[width=0.99\linewidth]{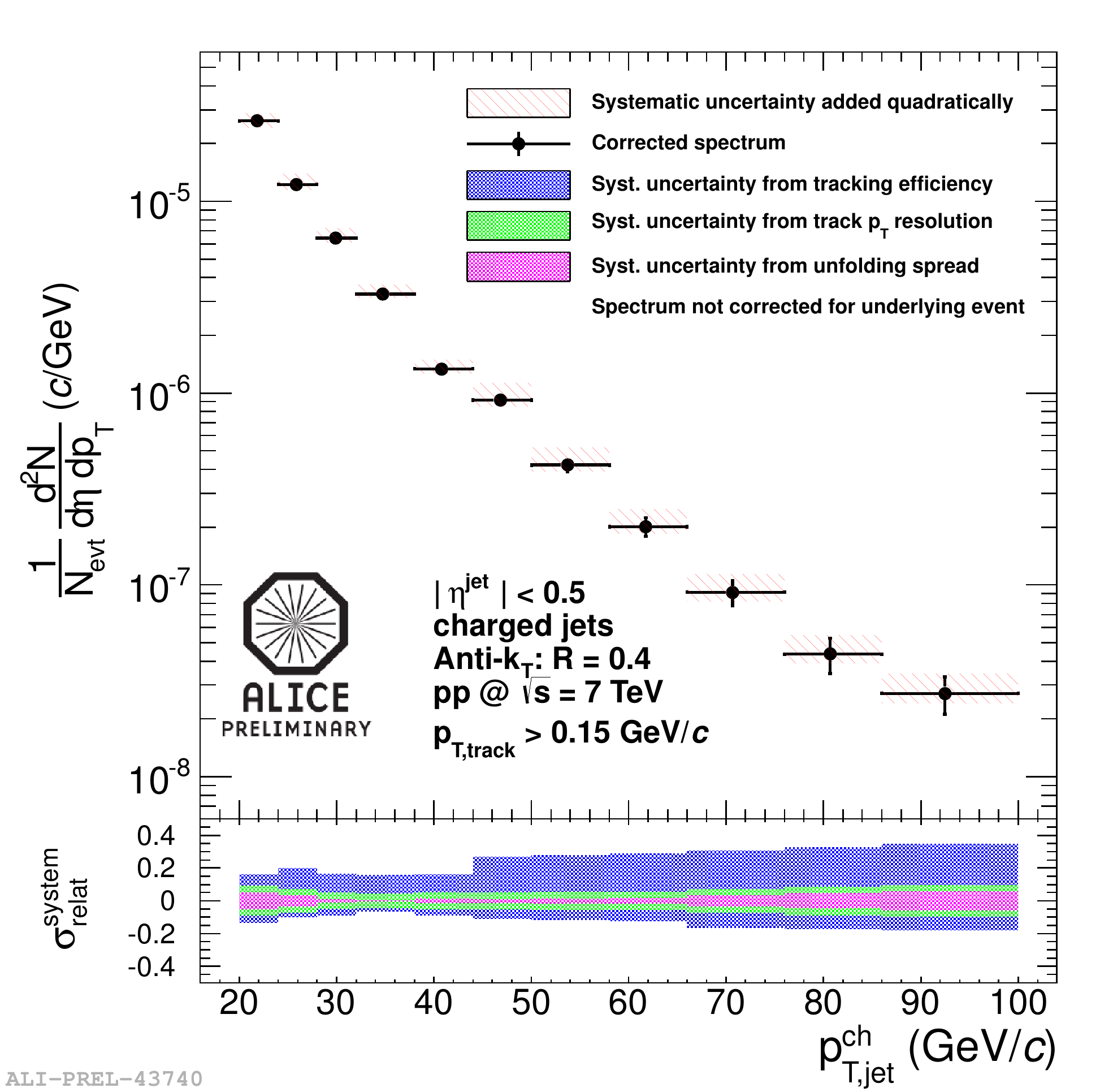}}
\end{minipage}

\begin{minipage}{0.37\linewidth}
\vspace{-0.1cm} \hspace{1.7cm}  \centerline{\includegraphics[width=1.06\linewidth]{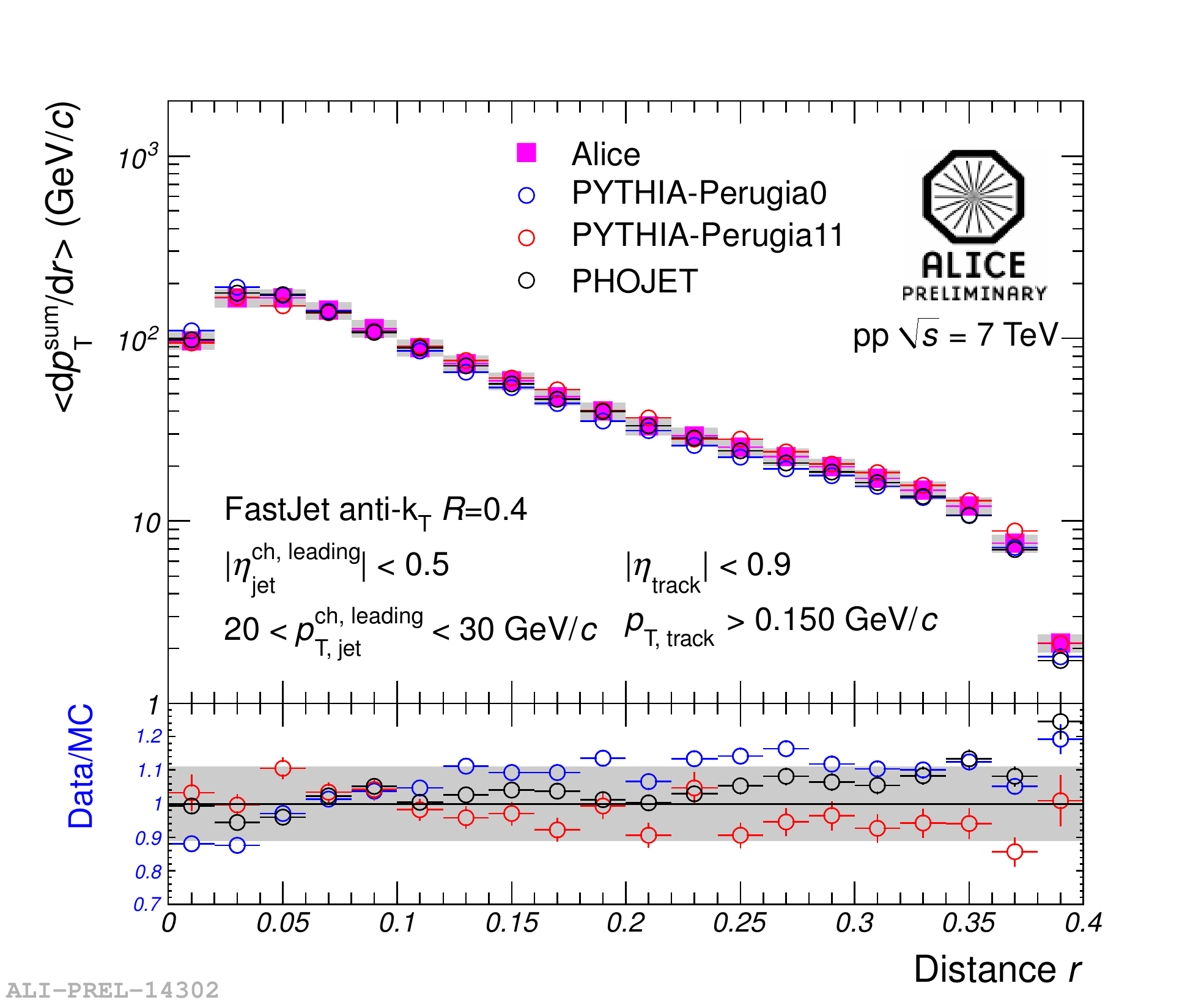}}
\end{minipage}
\begin{minipage}{0.37\linewidth}
\vspace{0.7cm} \hspace{1.8cm} \centerline{\includegraphics[width=1.04\linewidth]{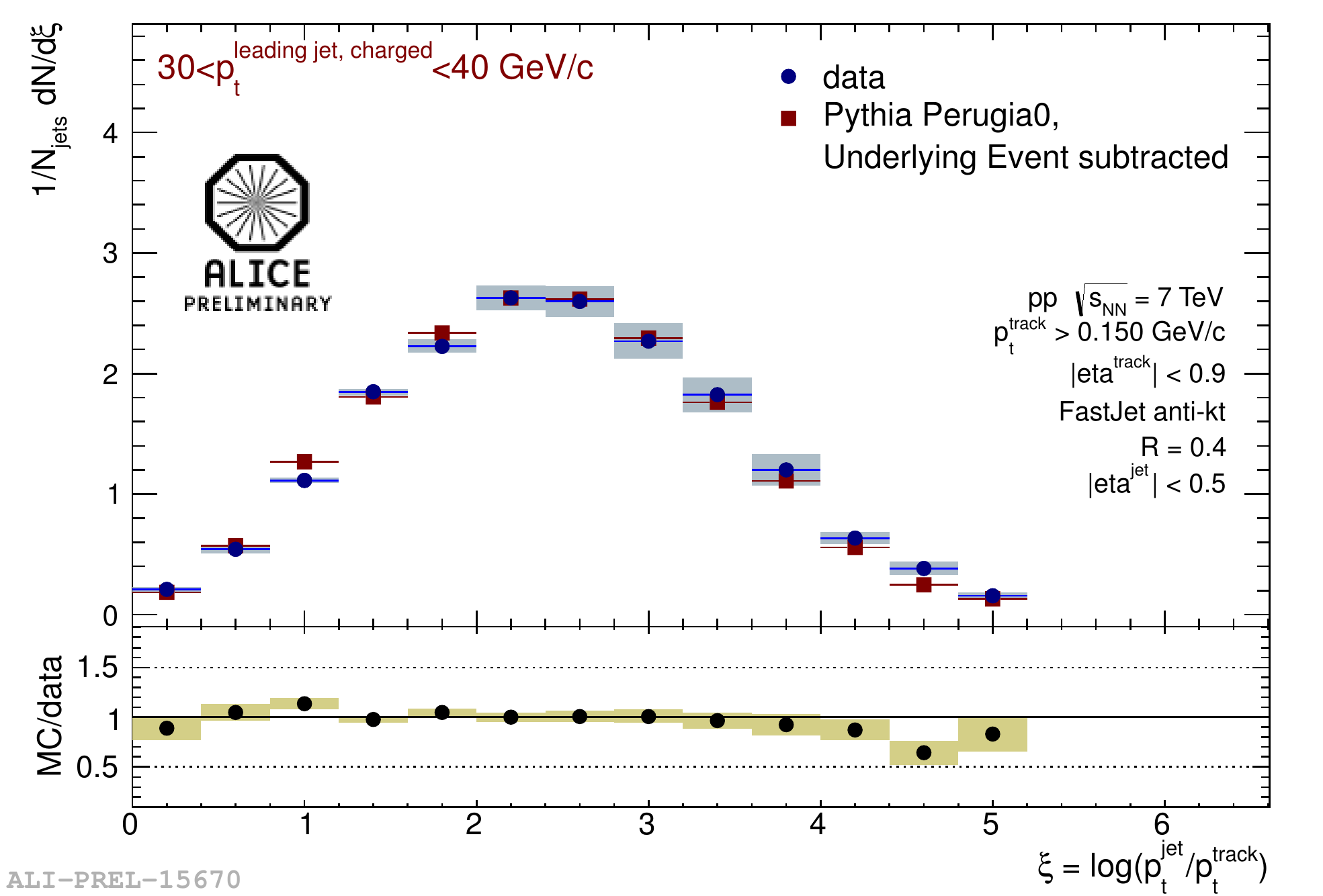}}
\end{minipage}
\caption{Results from pp collisions at $\sqrt{s}$=2.76~TeV and 7~TeV. Top left: inclusive differential jet cross section for 
$R$=0.4 at $\sqrt{s}$=2.76~TeV. Boxes show the systematic uncertainty. The data are compared to NLO pQCD calculations indicated by the 
bands, ratio of calculations 
to data are shown in the lower panel. Top right: charged jet cross section for $R$=0.4 at $\sqrt{s}$=7~TeV. Systematic 
uncertainties are indicated by the boxes, the lower panel shows the contributions from detector effects and unfolding superimposed linearly. 
Bottom left: radial distribution of transverse momentum density measured in leading charged jets 
with $20<p_{\rm T, jet}^{\rm ch}<30$~GeV/$c$. Bottom right: scaled transverse momentum distribution of jet fragments 
for $30<p_{\rm T, jet}^{\rm ch}<40$~GeV/$c$. Jet shape and fragmentation distributions are compared to from calculations from PYTHIA (tunes Perugia0 and Perugia2011) and Phojet MC event generators.}
\label{fig:pp}
\end{figure}

In the top-left panel of Fig.~\ref{fig:pp}, we present the inclusive differential jet cross sections~\cite{FullJetPaper} in pp 
collisions at $\sqrt{s}$=2.76~TeV, measured using charged+neutral particles. The data are compared to 
NLO pQCD results~\cite{NLO Armesto,NLO Soyez}: good agreement is found when 
hadronization effects are applied to the perturbative calculations. 

In the top-right panel of Fig.~\ref{fig:pp}, the charged jet cross section~\cite{Michal} at $\sqrt{s}$=7~TeV
is shown. The bottom-left panel of Fig.~\ref{fig:pp} presents the leading charged jet shape~\cite{Sidharth} distribution 
of particle $p_{\rm T}$ sum in radial slices relative to the jet axis. The slope of the distribution 
measures the jet collimation. The fragmentation spectrum of scaled transverse 
momentum $\xi = \ln \, (p_{\rm T}^{jet,ch}/p_{\rm T}^{particle})$ for the charged jet constituents~\cite{RongRong} shown in 
the bottom-right panel exhibits the characteristic hump-backed plateau structure indicating QCD coherence. The area of the 
distribution is equal to the charged particle multiplicity of the fragments. The measured jet shape and 
fragmentation are compared to MC simulations from the PYTHIA~\cite{PYTHIA} (tunes Perugia0 and Perugia2011) 
and Phojet~\cite{Phojet} event generators. The pp underlying event contribution was measured inside perpendicular cones transverse 
to the jet axis and subtracted from the jet cross section, fragmentation and shape measurements, consistently in data and simulations. 
The data are reasonably well described by the simulations.

\section{Jets in Pb-Pb Collisions}\label{sec:resultsPbPb}

\begin{figure}[h]

\begin{minipage}{0.40\linewidth}
\hspace{1.2cm} \centerline{\includegraphics[width=0.9\linewidth]{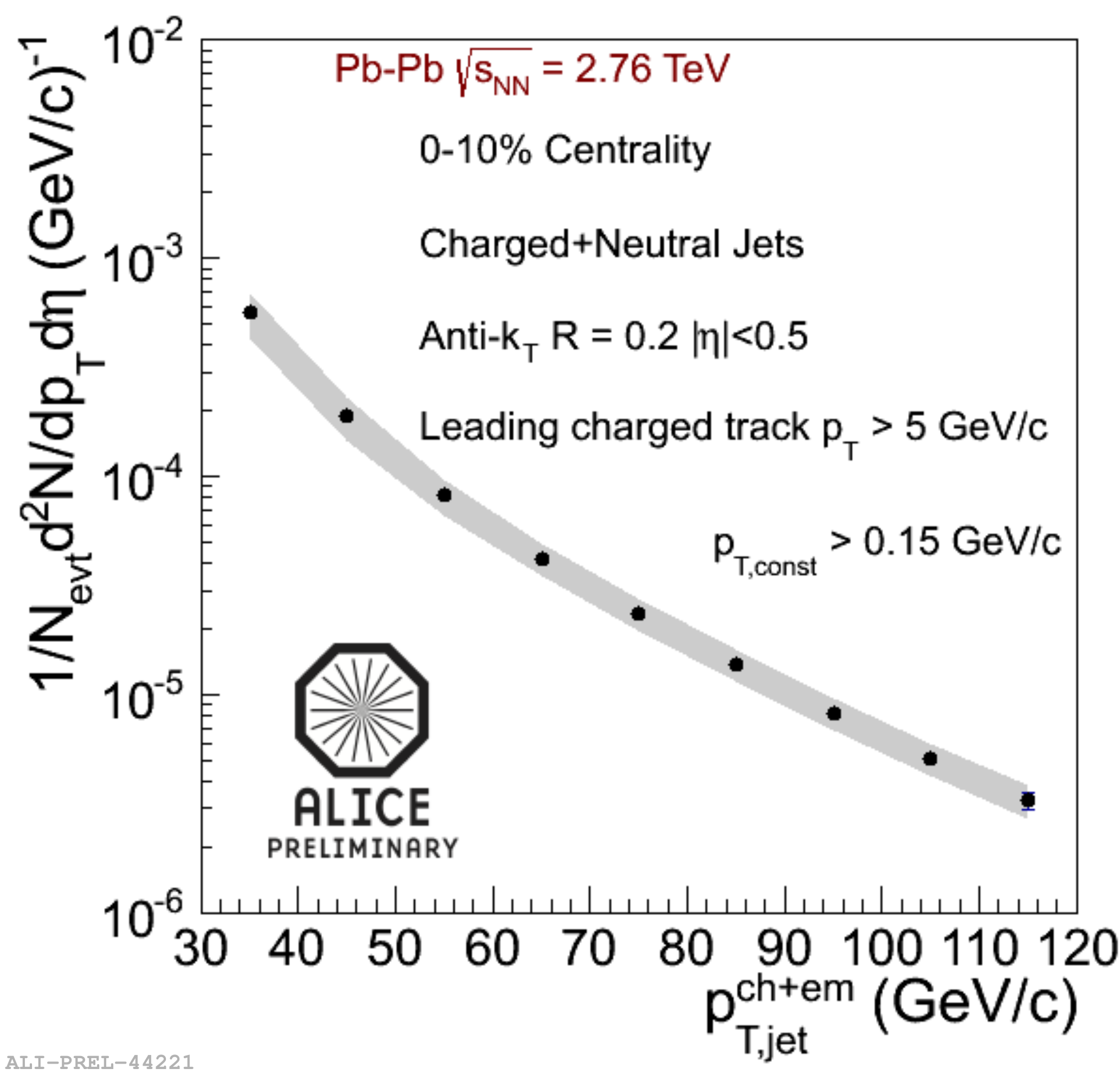}}
\end{minipage}
\begin{minipage}{0.40\linewidth}
\hspace{-2.5cm} \centerline{\includegraphics[width=0.9\linewidth]{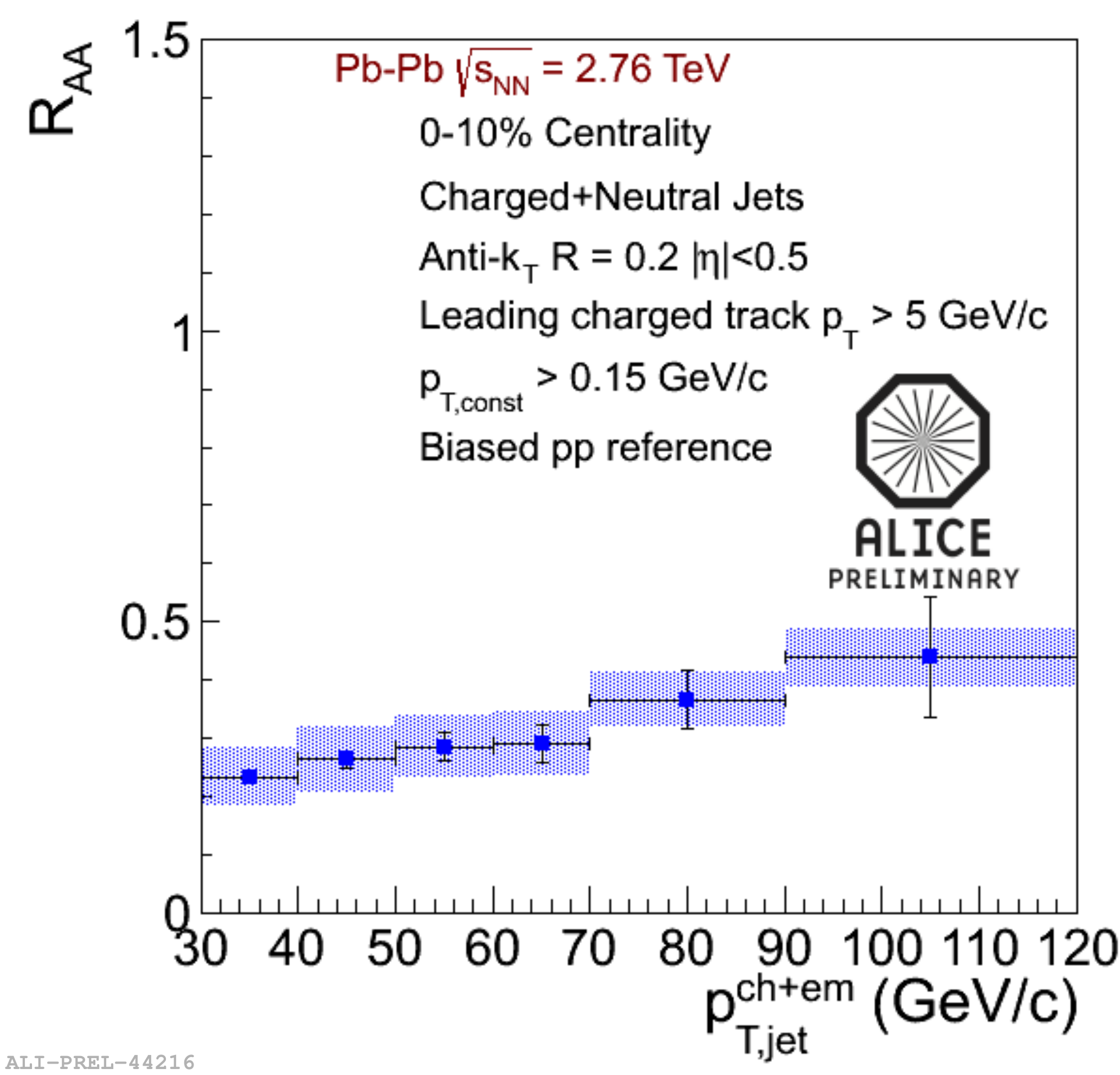}}
\end{minipage}
\begin{minipage}{0.40\linewidth}
\hspace{1.2cm}  \centerline{\includegraphics[width=0.99\linewidth]{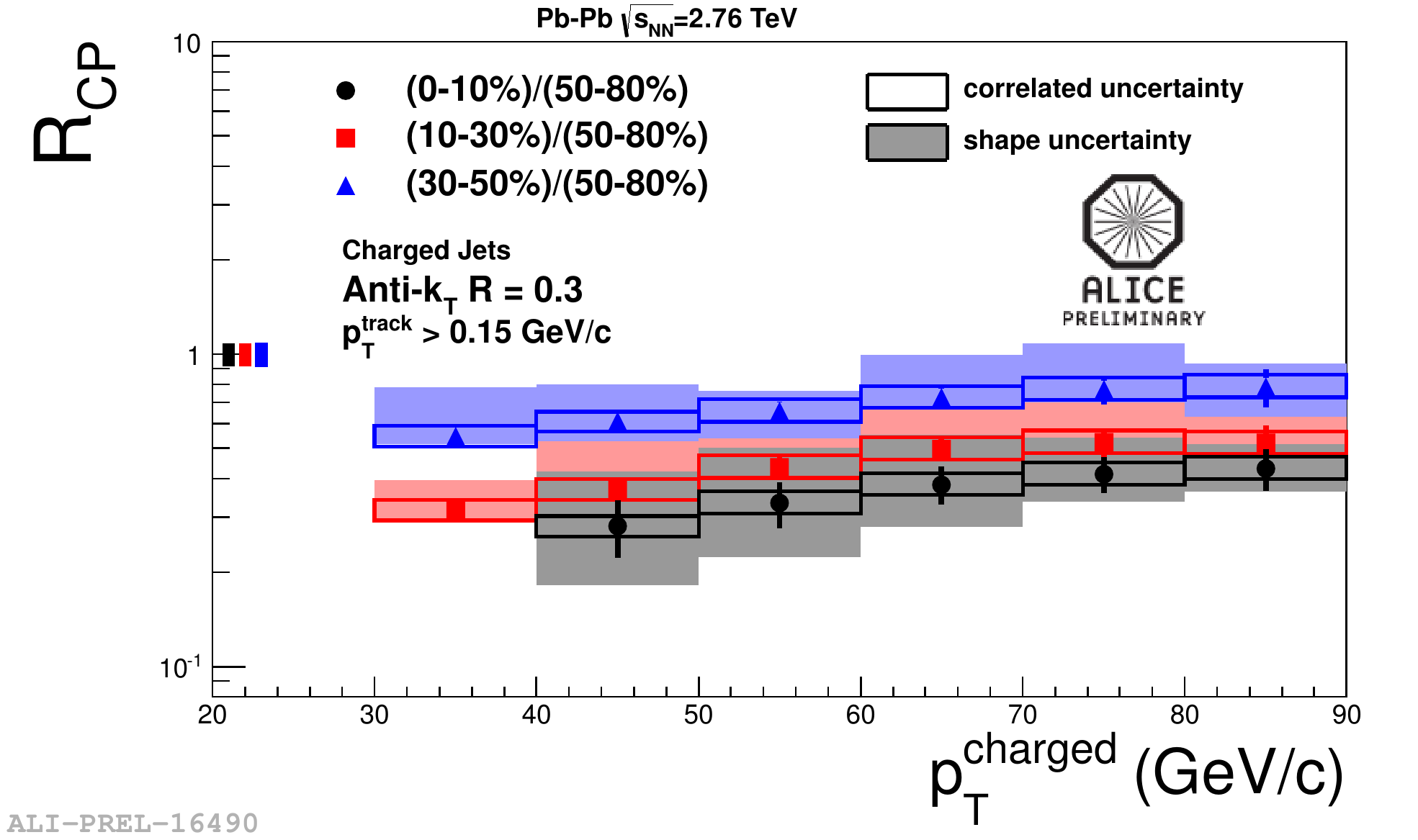}}
\end{minipage}
\begin{minipage}{0.40\linewidth}
\hspace{0.7cm} \vspace{-0.1cm} \centerline{\includegraphics[width=0.97\linewidth]{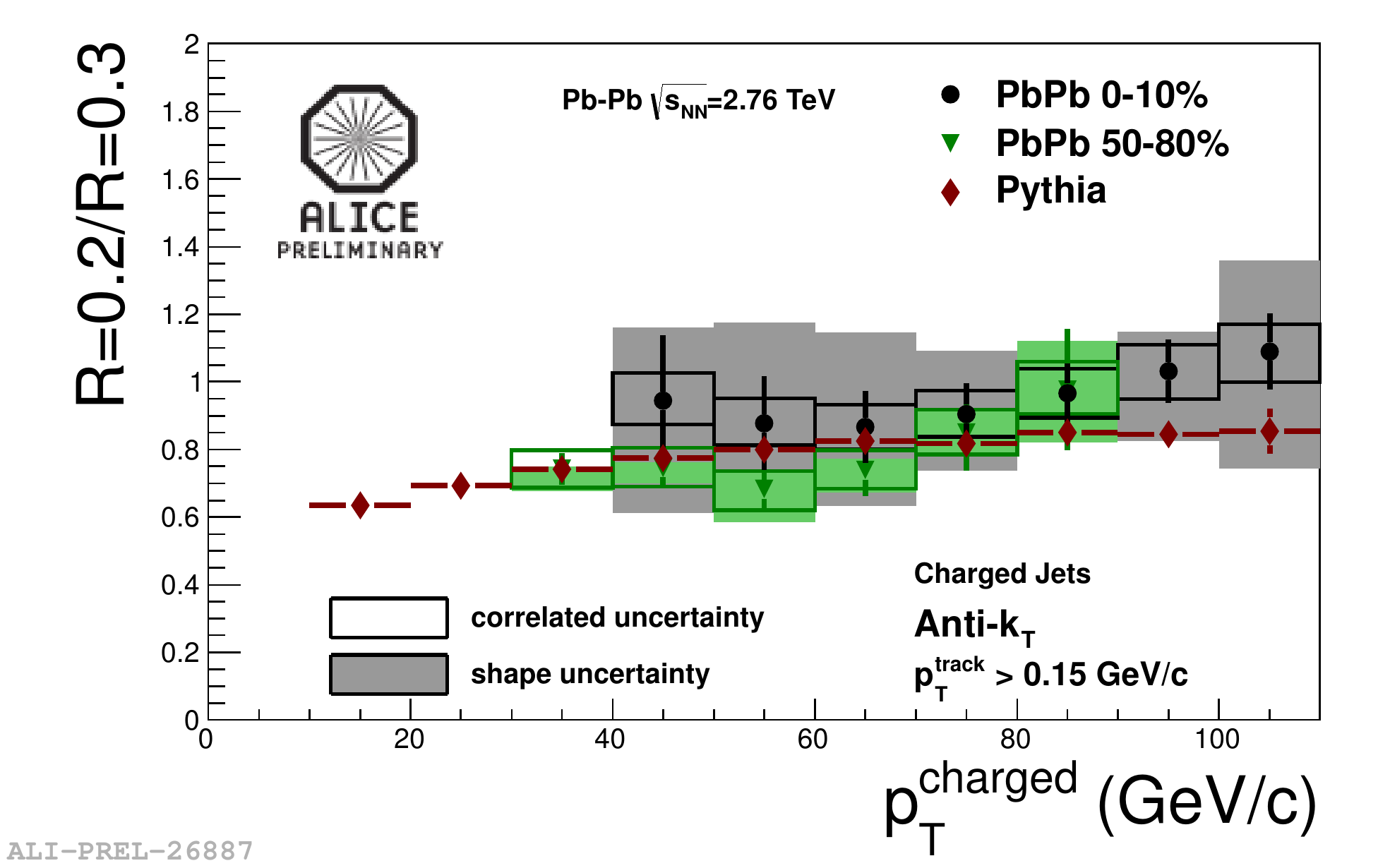}}
\end{minipage}

\caption{Single inclusive jet measurements in Pb-Pb collisions at $\sqrt{s}_{NN}=2.76$~TeV. Top left: inclusive jet spectrum for $R$=0.2 in the 
10\% most central events, with a 5 GeV/$c$ fragmentation bias on the leading jet constituent. Systematic uncertainties are indicated 
by the shaded band. Top right: nuclear modification factor for $R$=0.2 biased jets in 0-10\% central events. Bottom left: nuclear modification 
factor $R_{CP}$ for charged jets with $R$=0.3 for different centrality selections. Bottom right: ratio of charged jet spectra with radius 
parameter $R$=0.2 and $R$=0.3 in central and peripheral Pb-Pb events, compared to PYTHIA calculations.}
\label{fig:PbPb}
\end{figure}

Figure~\ref{fig:PbPb} shows the single inclusive jet cross section results~\cite{Rosi,Salvatore} in Pb-Pb 
collisions at $\sqrt{s}_{NN}=2.76$~TeV. In the top-left panel we present the charged+EMCal jet spectrum for $R$=0.2 for 
the 10$\%$ most central events. 
Minimum transverse momentum, $p_{\rm T}>5$~GeV, is required for the leading charged constituent. In the 
top-right panel the jet nuclear modification factor $R_{\rm AA}$ is shown, 
defined as the ratio of the measured jet spectra in Pb-Pb collisions to the pp reference scaled by the number of 
binary collisions~\cite{NColl}: $R_{\rm AA}(p_{\rm T}) = \frac{1} { \langle N_{\rm coll} \rangle} \frac {\mathrm{d}^2 N_{\rm jet} / \mathrm{d} p_{\rm T,jet} \; \mathrm{d}\eta} { \mathrm{d}^2 N_{\rm jet}^{\rm pp} / \mathrm{d} p_{\rm T,jet} \; \mathrm{d}\eta} $.  The measured $R_{AA}$ is significantly 
smaller than unity, indicating a strong suppression of jets in Pb-Pb collisions. The suppression is strongest at low jet 
momentum, at high momentum $R_{\rm AA}$ rises to values of the order of 0.5. The centrality 
dependence of the quenching is investigated via the charged jet $R_{\rm CP}$, a quantity conceptually similar 
to $R_{\rm AA}$, but using peripheral Pb-Pb events as a reference. $R_{\rm CP}$~\cite{Marta1,Marta2} is presented 
in Fig.~\ref{fig:PbPb} in the bottom-left panel for 3 different centrality intervals. The suppression is seen to 
increase with centrality (lower $R_{CP}$ values), and a more pronounced momentum dependence can be 
observed. In the bottom-right panel, we show, for two centrality classes, the ratio of the charged jet 
spectra obtained for $R$=0.2 to $R$=0.3. The ratio is sensitive to the jet structure. The measured 
ratio is consistent with the PYTHIA values for peripheral and central events: no broadening of 
the hard core of the reconstructed jets is observed within the present systematic uncertainties. 

\section{Hadron triggered recoil jets in Pb-Pb}\label{sec:PbPbRecoil}

The analysis of hadron triggered recoil jets allows us to extend the study of jet quenching to larger values of $R$.  
The semi-inclusive distribution of charged jets recoiling back-to-back from a high-pt charged hadron in Pb-Pb collisions~\cite{Leticia} is 
measured. Constructing the difference spectra for different intervals of trigger hadron $p_{\rm T}$, we subtract the 
contribution of combinatorial jets uncorrelated to the hadron trigger:

\[ \scriptstyle  \Delta_{recoil}((p_{\rm T}^{\rm trig,1} - p_{\rm T}^{\rm trig,2})-(p_{\rm T}^{\rm trig,3} - p_{\rm T}^{\rm trig,4})) 
= \frac{1}{N_{\rm trig}} \frac{\mathrm{d}N}{\mathrm{d} p^{\rm ch}_{\rm T,jet}}(p_{\rm T}^{\rm trig,1}<p_{\rm T}^{\rm trig}<p_{\rm T}^{\rm trig,2}) 
- c \frac{1}{N_{\rm trig}} \frac{\mathrm{d}N}{\mathrm{d} p^{\rm ch}_{\rm T,jet}}(p_{\rm T}^{\rm trig,3}<p_{\rm T}^{\rm trig}<p_{\rm T}^{\rm trig,4}) \]

\begin{figure}[h]
\begin{minipage}{0.38\linewidth}
\hspace{1.8cm} \centerline{\includegraphics[width=1.02\linewidth]{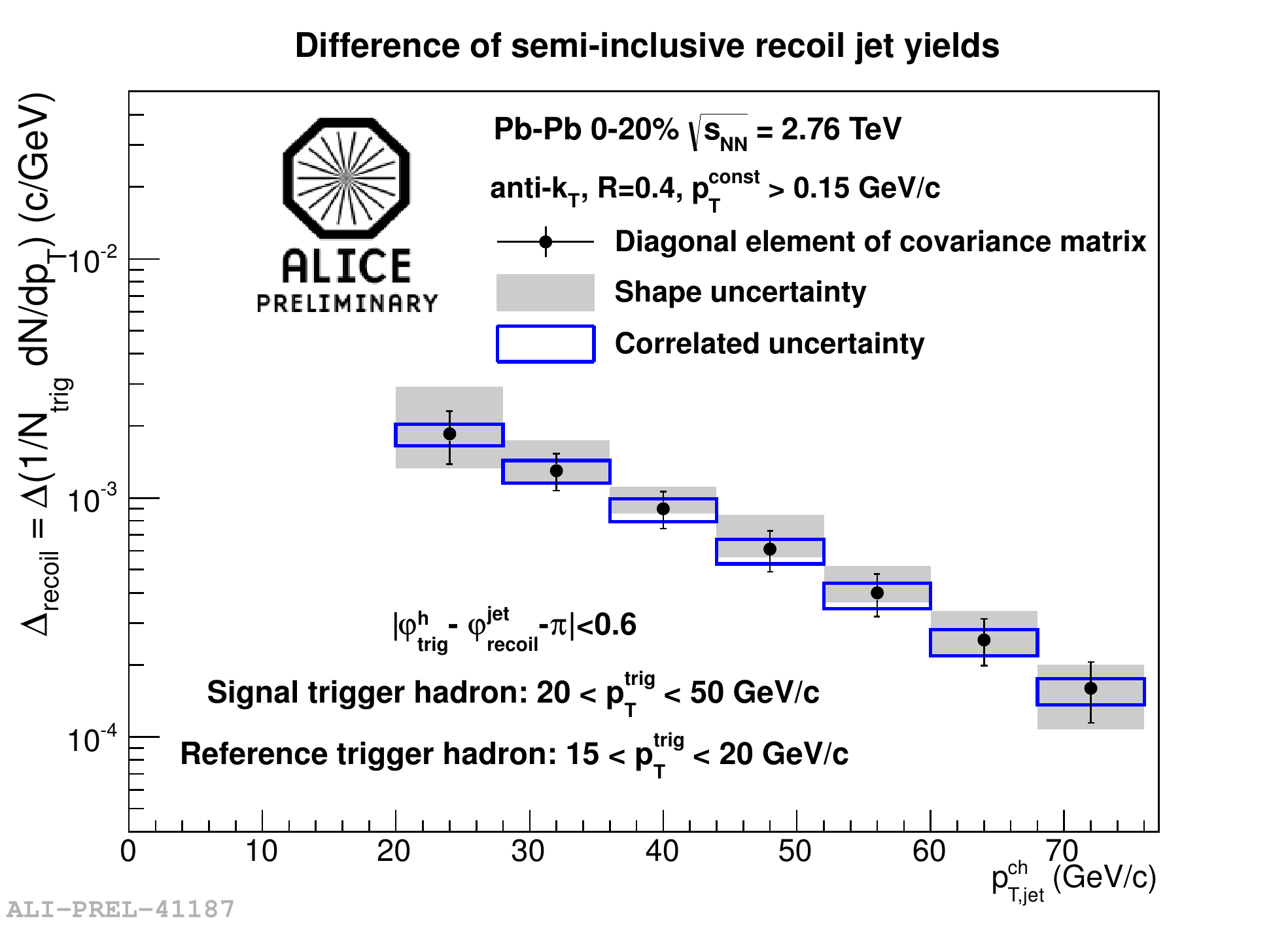}}
\end{minipage}
\begin{minipage}{0.38\linewidth}
\hspace{1.8cm} \centerline{\includegraphics[width=1.02\linewidth]{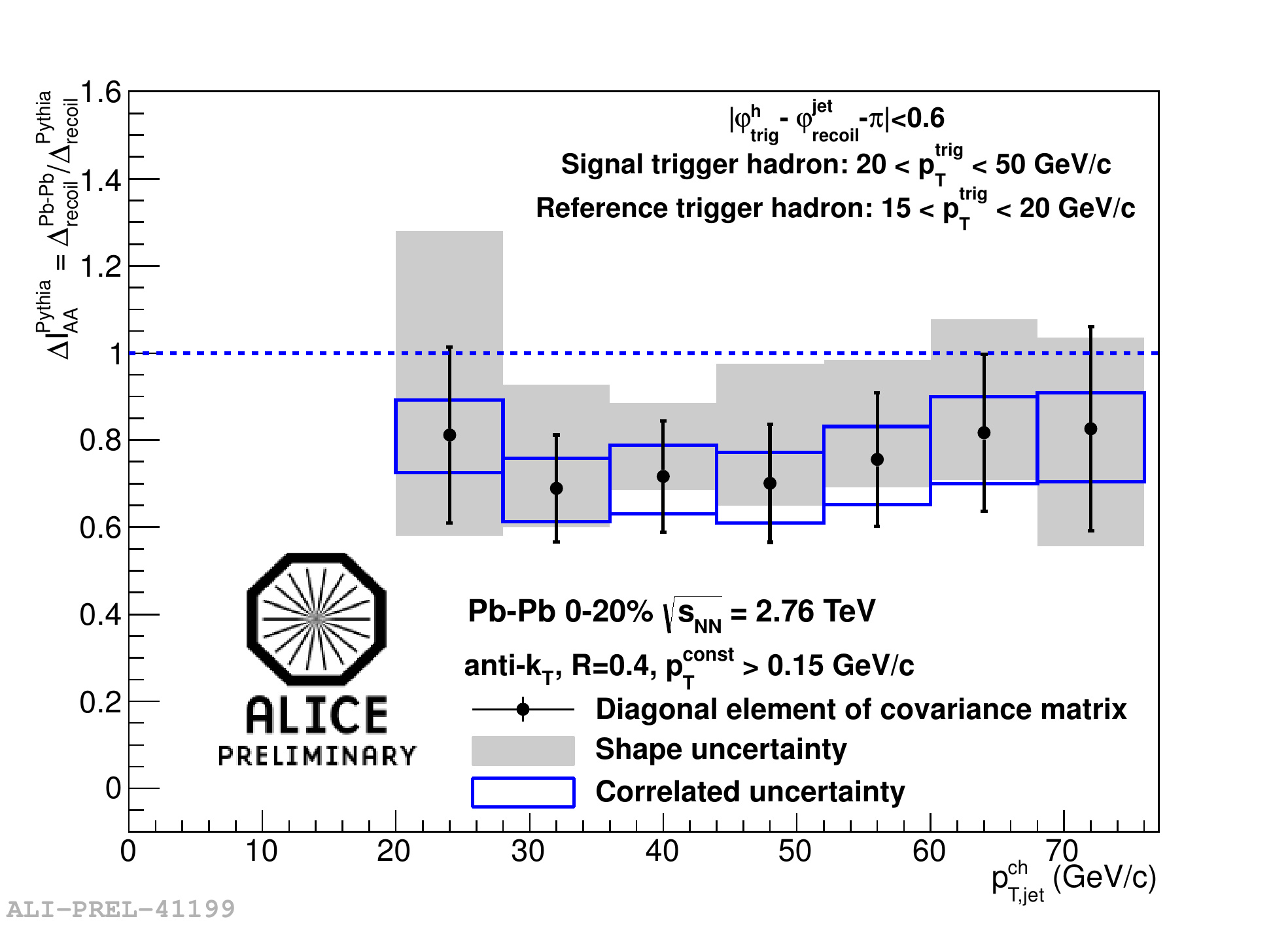}}
\end{minipage}
\caption{Semi-inclusive recoil jet spectrum difference $\Delta_{recoil}$ for 0-20\% central collisions, $p_{\rm T}^{const}>0.15$~GeV/$c$, 
         jets reconstructed with anti-$k_{\rm T}$ for $R$=0.4 (left). $\Delta_{IAA}$  for recoil jets using PYTHIA as a reference (right).}
\label{fig:recoil}
\end{figure}

The $\Delta_{\rm recoil}$ spectrum for $R$=0.4 is shown in the left panel of Fig.~\ref{fig:recoil} for hadron trigger 
intervals from 20-50~GeV/$c$ and 15-20~GeV/$c$, using a scale factor $c = 0.956$. This novel variable 
allows us to study an unbiased sample of jets reconstructed with a large radius, maintaining a jet constituent 
cutoff as low as $p_{\rm T}^{\rm const}>0.15~$GeV/$c$. \par
To explore the energy redistribution within recoil jets, we consider the ratio for the measured $\Delta_{recoil}$ distribution 
over PYTHIA (tune Perugia 2010) simulations $\Delta I_{AA}^{PYTHIA}$, presented in the right panel of Fig.~\ref{fig:recoil}. We 
find consistent $\Delta I_{AA}^{PYTHIA}$ for different values of $R$ and constituent $p_{\rm T}$ 
threshold~\cite{Leticia} (not shown). Within  present experimental uncertainties, we do not observe 
significant redistribution of the jet energy.

\end{document}